# Angular dependence of the fluctuation magnetization vector above the superconducting transition of a highly anisotropic high-$T_c$ cuprate

R.I. Rey[1], J. Mosqueira[1,*], N. Cotón[1], J.D. Dancausa[1], J.M. Doval[1], A. Ramos-Álvarez[1], A. Wahl[2], M. Tello[3], F. Vidal[1]

[1]*LBTS, Universidade de Santiago de Compostela, ES-15782 Santiago de Compostela, Spain*
[2]*Laboratoire CRISMAT, ENSICAEN, UMR 6508 CNRS, Caen, France*
[3]*UPV, Departamento de Física de la Materia Condensada,Bilbao, Spain*
*j.mosqueira@usc.es, Tel.: +34-981-563100, Fax: +34-881-814 112

**Abstract:**

In highly anisotropic cuprate superconductors it is generally accepted that the reversible magnetization vector, $\vec{M}$, is essentially perpendicular to the superconducting $CuO_2$ layers in a wide range of crystal orientations with respect to the applied magnetic field, $\vec{H}$. In a recent work [J. Mosqueira et al., Phys. Rev. B **84**, 134504 (2011)] it is shown, however, that the dependence of $\vec{M}$ on the $\vec{H}$ orientation in the reversible mixed state of a high-quality Tl-based cuprate presents a notable deviation from this behavior. Here we extend these measurements to the fluctuation region above $T_c$, in order to check whether the above mentioned effect is also present.

1. Introduction

A central result when describing the reversible magnetic properties of highly anisotropic cuprate superconductors is that the magnetization vector in *tilted* crystals is perpendicular to the crystal *c*-axis in a wide range of orientations, and is only dependent on the component of the applied field along it [1], i.e.,

$$M_\perp(H,\theta) = M_\perp(H\cos\theta,0), \qquad (1)$$

and

$$M_\parallel \approx 0. \qquad (2)$$

Here $M_\perp$ and $M_{//}$ are the $\vec{M}$ components perpendicular and, respectively, parallel to the CuO$_2$ layers, and $\theta$ the angle between $\vec{H}$ and the crystal $c$ axis. Eqs. (1) and (2) are commonly used to interpret measurements of the magnetic torque in high-$T_c$ cuprates (for a recent example see Ref. 2). In a recent work, however, notable deviations with respect to this behaviour were reported [3]. In particular, by using a high-quality Tl-based highly-anisotropic single crystal, the magnetization vector in the reversible London region was shown to present a significant angular slippage from the crystal $c$ axis when this last is tilted from the applied field direction. Lawrence-Doniach approaches for single-layered highly anisotropic superconductors do not account for such an observation [4], and it was then proposed that it could be due to the multilayered nature of the compound studied. Here we extend the study of Ref. 3 to the region above $T_c$ in which fluctuation effects are relevant, in order to check whether a similar breakdown of Eqs. (1) and (2) is also present.

**2. Experimental details and results**

The Tl-2223 sample used in this work is a plate-like single crystal (1.1x0.75x0.226 mm$^3$) with the $c$ crystallographic axis perpendicular to the largest face. Details of its growth procedure may be seen in Ref. 5. Let us just mention that this crystal was already used in the magnetization measurements presented in Refs. 3,6,7, and it has a sharp low-field diamagnetic transition ($T_c$ = 122 ± 1 K), a large Meissner fraction (~ 80%), an excellent crystallinity (its mosaic spread is as low as ~0.1°), and a extreme anisotropy (the anisotropy factor is at least γ ~ 200). [6,7] The magnetization measurements were performed with a Quantum Design SQUID magnetometer equipped with independent detectors for the components of the magnetic moment in the direction of the applied magnetic field (hereafter *longitudinal*) and in a direction *transverse* to it ($m_L$ and $m_T$ respectively).

The sample was glued with GE varnish to a sample holder (also from Quantum Design) which allows rotations about an axis perpendicular to the field direction. The orientation may be specified with a precision of 0.1° with a reproducibility of ±1°. A schematic diagram of the experimetal set-up is presented in Fig. 1, where the choice for the ($\perp$, $\parallel$) and ($L,T$) axes is also indicated. $m_L$ and $m_T$ were measured against temperature by using different constant $\theta$ and $H$ values. The measurements range from temperatures below $T_c$ up to ~250 K (~$2T_c$), which allowed to characterize with accuracy the *background* contribution to the magnetic moment, mainly coming from the rotating sample holder. The fluctuation contribution to the magnetization was then obtained by subtracting this background contribution to the raw data (some examples of this procedure are presented in detail in Ref. 3). The resulting longitudinal and transverse fluctuation-induced magnetic susceptibilities are presented in Figs. 2 and 3 as a function of temperature and for different magnetic field amplitudes and orientations.

## 3. Data analysis

According to Fig. 1, and taking into account Eqs. (1) and (2) for highly anisotropic superconductors, the component of the magnetization vector in the direction of the applied magnetic field may be approximated by

$$M_L(T, H, \theta) \approx M_\perp(T, H \cos\theta) \cos\theta, \qquad (3)$$

and the component transverse to the field

$$M_T(T, H, \theta) \approx -M_\perp(T, H \cos\theta) \sin\theta. \qquad (4)$$

In the framework of the Gaussian Ginzburg-Landau approach, when $H \perp ab$ the magnetization of highly anisotropic superconductors induced above $T_c$ by thermal fluctuations may be expressed as [8]

$$M_\perp(T, H, \theta = 0) = -f \frac{k_B T N}{\phi_0 s} \left[ \frac{c-\varepsilon}{2h} + \frac{\varepsilon}{2h} \psi\left(\frac{h+\varepsilon}{2h}\right) - \frac{c}{2h} \psi\left(\frac{h+c}{2h}\right) - \ln\Gamma\left(\frac{h+\varepsilon}{2h}\right) + \ln\Gamma\left(\frac{h+c}{2h}\right) \right]. \qquad (5)$$

Here $\Gamma$ and $\psi$ are the gamma and digamma functions, $h \equiv H/H_{c2}^\perp(0)$ the reduced magnetic field, $H_{c2}^\perp(0)$ the upper critical field extrapolated to $T = 0$ K, $\varepsilon \equiv \ln(T/T_c)$ the reduced temperature, $N = 3$ the number of superconducting $CuO_2$ layers in their periodicity length ($s = 1.79$ nm), $f \approx 0.8$ the effective superconducting volume fraction (approximated as the Meissner fraction), $k_B$ the Boltzmann constant, $\phi_0$ the flux quantum, and $c \approx 0.55$ the *total-energy cutoff* constant [9]. This expression is valid in the so-called finite-field or Prange regime. However, as $\mu_0 H_{c2}^{\perp}(0) \approx 300$ T [6], even for the largest fields used in the experiments (2 T) the experimental data are in the low field limit ($h \ll \varepsilon$) except for a narrow temperature region just above $T_c$. In this limit, Eq. (1) may be simplified to

$$M_{\perp}(T, H, \theta = 0) = -f \frac{k_B T N \pi \mu_0 H \xi_{ab}^2(0)}{3\phi_0^2 s} \left( \frac{1}{\varepsilon} - \frac{1}{c} \right), \tag{6}$$

which is linear in *H*. It is worth noting that in absence of cutoff (i.e., when $c \to \infty$), Eq. (6) reduces to the two-dimensional version of the well known Schmidt expression for the magnetization induced by superconducting fluctuations above $T_c$ [10]. By combining Eqs. (6), (3) and (4) it finally results

$$M_L(T, H, \theta) = -f \frac{k_B T N \pi \mu_0 H \xi_{ab}^2(0)}{3\phi_0^2 s} \left[ \frac{1}{\ln(T/T_c)} - \frac{1}{c} \right] \cos^2 \theta, \tag{7}$$

and

$$M_T(T, H, \theta) = f \frac{k_B T N \pi \mu_0 H \xi_{ab}^2(0)}{3\phi_0^2 s} \left[ \frac{1}{\ln(T/T_c)} - \frac{1}{c} \right] \sin \theta \cos \theta. \tag{8}$$

The lines in Figs. 2 and 3 correspond, respectively, to Eqs. (7) and (8). They were evaluated by using with $\xi_{ab}(0) = 1.0$ nm, which is the value leading to a good agreement with the data corresponding to $H \perp ab$ (see Ref. 6). Taking into account the uncertainty in the determination of the background signal (mainly due to the rotating sample holder), the agreement with the experimental data under the lowest applied fields is reasonably good down to a few degrees above $T_c$, where the Gaussian approximation is no longer valid. On the one side, this result further confirms the adequacy of Ginzburg-Landau approaches to describe fluctuation effects in high-$T_c$ cuprates, at present a highly debated issue [2]. On the other, it suggests that in the fluctuation region above $T_c$, Eqs. (1) and (2) are applicable, which means that the diamagnetic behavior observed above $T_c$ in tilted crystals results from currents, created by the thermally induced Cooper pairs, *confined* in the $CuO_2$ layers, the possible interlayer currents playing a negligible role.

**Acknowledgments**

This work was supported by the Spanish MICINN and ERDF (FIS2010-19807), the Xunta de Galicia (2010/XA043 and 10TMT206012PR), and the european project ENERMAT. We acknowledge J. Ponte for his valuable help with the rotating sample holder.

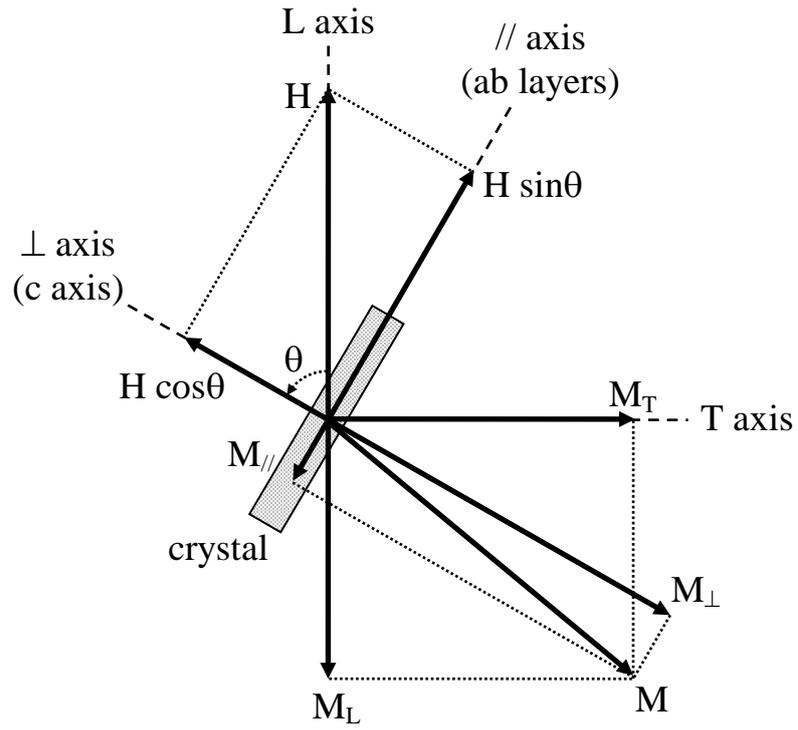

**Fig. 1.** Schematic diagram indicating the choice for the ($\perp$,//) and (L,T) axes, the corresponding components of the magnetization and magnetic field vectors, and the definition of $\theta$.

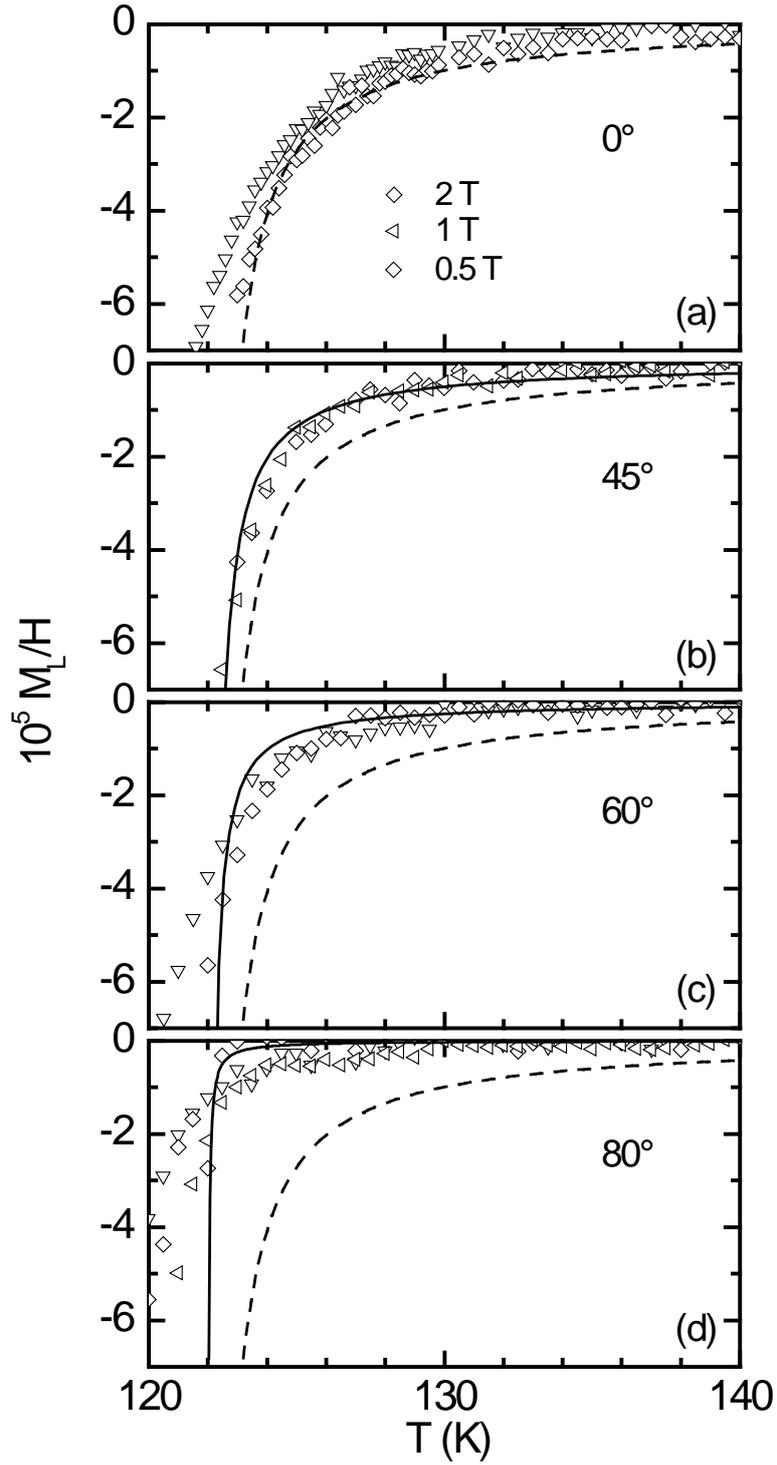

**Fig. 2.** Temperature dependence of the *longitudinal* (in the direction of the applied magnetic field) fluctuation magnetic susceptibility above $T_c$ for different magnetic field amplitudes and orientations. The spread below $T_c$ is due to the breakdown of the Gaussian approximation. The lines correspond to Eq. (7) evaluated by using $\xi_{ab}(0) = 1.0$ nm. To better appreciate the reduction in the $\Delta M_L/H$ amplitude on increasing $\theta$, the dashed curve in (a), corresponding to $\theta = 0°$, is also included in (b), (c) and (d).

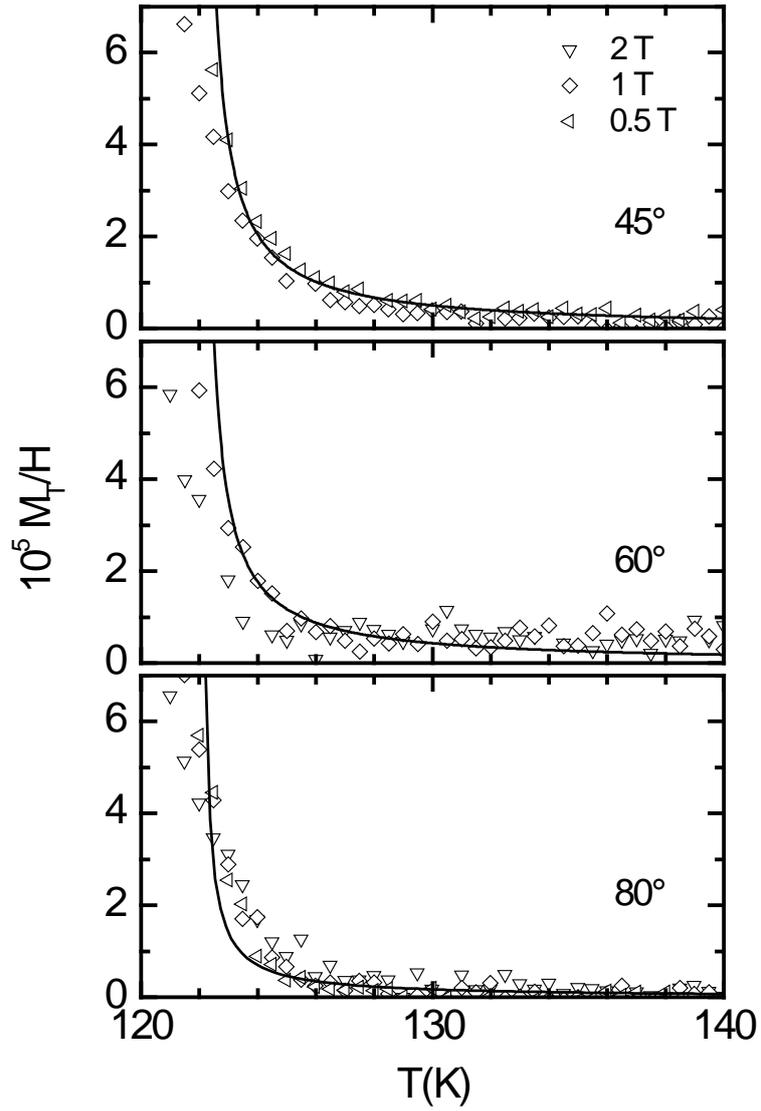

**Fig. 3**. Temperature dependence of the *transverse* (in the direction perpendicular to the applied magnetic field) fluctuation magnetic susceptibility above $T_c$ for different magnetic field amplitudes and orientations. The spread below $T_c$ is due to the breakdown of the Gaussian approximation. The lines correspond to Eq. (8) evaluated by using $\xi_{ab}(0) = 1.0$ nm.